\documentclass[prl,twocolumn,twoside,floatfix,showpacs]{revtex4}
\usepackage{graphicx,amsmath,amssymb,mathptm}

\begin{document}

\title{Flow correlated percolation during vascular network formation 
in tumors}

\author{D.-S. Lee and H. Rieger}
\affiliation{Theoretische Physik, Universit\"{a}t des Saarlandes, 
66041 Saarbr\"{u}cken, Germany}

\author{K. Bartha}
\affiliation{Department of Medical Biochemistry, Semmelweis University, 
Budapest, Hungary}

\date{\today}

\begin{abstract}
A theoretical model based on the molecular interactions between a
growing tumor and a dynamically evolving blood vessel network
describes the transformation of the regular vasculature in normal
tissues into a highly inhomogeneous tumor specific capillary network.
The emerging morphology, characterized by the compartmentalization of
the tumor into several regions differing in vessel density, diameter
and necrosis, is in accordance with experimental data for human
melanoma. Vessel collapse due to a combination of severely reduced
blood flow and solid stress exerted by the tumor, leads to a
correlated percolation process that is driven towards 
criticality by the mechanism of hydrodynamic 
vessel stabilization.
\end{abstract}
\maketitle

Tumor vasculature, the network of blood vessels in and around a
growing tumor, is in many respects different from the regular
vasculature in normal tissues. Hypoxia, the lack of oxygen, that
prevents a small tumor nucleus from further growth, induces the
expression of various diffusible growth factors (GF) by the tumor
cells that trigger a coordinated response of angiogenesis - the
formation of irregular blood vessels (for a review see
\cite{carmeliet-jain,acker-plate}). The expected increase in 
microvasular density (MVD) is usually observed in the periphery of the
tumor, whereas the morphology of the vasculature in the central part
of the tumor is characterized by a {\it decreased} MVD, dilated vessels
and regions of necrotic tumor tissue \cite{holash,paku}. The resulting
tumor specific capillary network is very heterogeneous, composed of
dense and void regions, and has a fractal dimension different from
normal arteriovenous or normal capillary networks
\cite{jain-fractal}.

Although on the molecular level the main actors in the angiogenic game
are rapidly identified, the physical principles that determine the
global morphology of the vascular network in tumor tissues are not
known. Since for instance MVD is used as a diagnostic tool in cancer
therapy \cite{mvd-review} a quantitative understanding of the
mechanism that leads to the compartmentalization of the tumor
vasculature into various regions differing substantially in vessel
density appears mandatory. Moreover, scale-invariant aspects like
fractal dimension, are used as hints towards the nature of the growth
process underlying the formation of the tumor vasculature
\cite{fractal-review}. In this Letter we propose a theoretical model
for the evolution of tumor vasculature that illuminates the physical 
principles leading to its global morphology. The experimentally observed
increase in MVD at the tumor perimeter and periphery and decrease 
in MVD and vessel dilation in the tumor center in human melanoma 
\cite{paku} appears also as the general scenario in the theoretical model 
that we discuss. Furthermore, we will argue that vessel collapses in the
interior of the tumor lead to a percolation process which is driven
towards criticality, the percolation threshold, via a mechanism of
vessel stabilization by increased blood flow in the remaining vessels.

Guided by a two-dimensional cellular automaton model that two of us
developed recently \cite{2d-model} we consider the tumor-vessel system
as a dynamically evolving network or graph interacting with a tumor
growth process (inspired by the Eden model \cite{eden}). The
interaction takes place via two concentration fields: the oxygen
originating in the vessel network, and the growth factor originating
in the tumor cells (TC). A hydrodynamic flow is imprinted on the
vessel network that emits oxygen. TC's proliferate/die when the local
oxygen concentration is high/low.  Vessels (edges) emerge when the
local GF concentration is high enough, and they vanish (collapse)
stochastically inside the tumor, when the hydrodynamic shear force
acting on the vessel walls is too low. The biological and
pathophysiological motivation for the details of the model definition
to follow is discussed in
\cite{2d-model}.

To be specific, we describe the topology of the vessel network by a
graph $G=(V,E)$, and identify each edge $e\in E$ with a vessel, and
each node $v\in V$ with a vessel junction, where more than two vessels
merge. Here we restrict to capillary networks and do not discriminate
between arteries and veins. The network $G$ is embedded in the
three-dimensional Euclidean space $R^3$ and restricted to the cube
$Z\subset R^3$ of volume $\mathcal{L}^3$. This cube is discretized into
$L^3=(\mathcal{L}/a)^3$ unit cells, where a dimensionless vector ${\bf
r}=x{\bf i}+y{\bf j}+z{\bf k}$ with $x,y,z=0,1,2,\ldots,L-1$ denotes
each unit cell. The microscopic length scale is chosen to be 
$a=10\mu m$, the typical size of the endothelial cell (EC) and TC.

For computational convenience we restrict the edges to run only
parallel to the three coordinate axes and identify an edge with the
string of unit cells of $Z$ that it covers: Let ${\bf r}_s(e)$ and
${\bf r}_t(e)$ be the two end-points of an edge $e\in E$
and $\ell(e)=|{\bf r}_t(e)-{\bf r}_s(e)| =n(e)-1$ 
the length of the vessel, then
\begin{equation}
e=\{ {\bf r}={\bf r}({\bf r}_s(e),{\bf r}_t(e),\zeta)\;|\;
\zeta=0,1,2,\ldots,\ell(e)\}
\end{equation}
with ${\bf r}({\bf r}_s,{\bf r}_t,\zeta) \equiv {\bf r}_s + 
\zeta ({\bf r}_t-{\bf r}_s)/\ell(e)$. 
Note that $V=\{{\bf r}_s(e)|e\in E\}\cup\{{\bf r}_t(e)|e\in
E\}$. 

The tumor is represented by the set $T$ of points that are occupied by
tumor cells: $T=\{{\bf r}|{\rm A\ TC\ exists \ at\ } {\bf r}\}$. The
vessel network $G$ is the source of an oxygen concentration field
$O_2({\bf r})$ and the tumor $T$ is the source of a growth factor
concentration field $GF({\bf r})$:
\begin{eqnarray*}
GF({\bf r}) &=& \sum_{{\bf r}'\in T} 
h_{R_{\rm gf}}(|{\bf r}-{\bf r}'|), \nonumber\\
O_2({\bf r}) &=& \sum_{{\bf r}'\in E} 
h_{R_{\rm oxy}}(|{\bf r}-{\bf r}'|).
\end{eqnarray*}
$R_{\rm gf}$ and $R_{\rm oxy}$ are the growth factor and oxygen
diffusion radii, respectively, and for simplicity we choose a
piecewise linear and normalized form for the contribution $h_R(r)$ of
each tumor cell / vessel segment, $h_R(r)=(1-r/R)/(\pi R^3/3)$ for
$r<R$ and $h_R(r)=0$ for $r\geq R$, satisfying $\int_0^\infty dr \,
h_R(r) 4\pi R^2 =1$.

Each edge $e$ represents a tubular vessel of diameter $d(e)$, through
which a hydrodynamic blood flow of magnitude $q(e)$, exerting a shear
force $f(e)$ upon the vessel walls, can pass. The flow is assumed to 
be an incompressible laminar stationary flow for which $q(e)$ and
$f(e)$ are calculated by Poiseuille's law:
\begin{equation}
q(e)=d^4(e) \nabla P(e), \quad {\rm and} \quad f(e)=d(e) \nabla P(e),
\label{eq:poiseuille}
\end{equation}
where the pressure gradient in the vessel $e$ is defined via $\nabla
P(e) =|P({\bf r}_t(e))-P({\bf r}_s(e))|$ and the pressure $P({\bf r})$
in the nodes (vessel junctions) of the network is computed using
Kirchhoff's law. The boundary condition for the blood
pressure $P({\bf r})$ is static and chosen in such a way that blood
flow as well as shear force in the original network is homogeneous:
$P({\bf r}=(x,y,z)) = 1.5 |x+y+z|/[3(L-1)]$ at the boundary $\partial
Z=\{{\bf r}=(x,y,z)|x=0\ {\rm or} \ y=0 \ {\rm or}\ z=0\}$.

\underline{\it Initial configuration:} The original tissue is regularly
vascularized with a homogeneous capillary network of given MVD that 
is fixed by inter-vessel distance $\delta$: $E=E_0=\{e|{\bf r}_s(e) =
\delta(n_1,n_2,n_3), {\bf r}_t(e)={\bf r}_s(e)+\delta {\bf i},\delta
{\bf j}, \ {\rm or}\ \delta{\bf k}, \ n_1,n_2,n_3=0,1,\ldots, N-1\}$ 
with $N=\lfloor L/\delta \rfloor+1$.  The number of nodes $n(V)$ is
$N^3$, that of edges $n(E)$ is $3(N-1)N^2$, and $n(e)=\delta+1$. For
each edge $e$, $d(e)=1(10\mu m)$, $q(e)=q_0$, and $f(e)=f_0$ with
$q_0=f_0=0.5/L$ from Eq.~(\ref{eq:poiseuille}). A tumor nucleus
containing $N_{TC}$ tumor cells and grown using the Eden rule
\cite{eden} starting with a seed at the system center ${\bf
r}_c=(L/2,L/2,L/2)$ defines the set of tumor cells $T$ at time
$t=0$.  $t_{\rm uo}({\bf r})=0$ for all ${\bf r}\in T$, 
representing the time spent in hypoxia.  Starting with
this initial configuration the following computations are performed
sequentially in each time step of duration $\Delta t = 1h$.

\underline{\it TC proliferation:} Proliferation of tumor cells is 
possible at tumor surface sites ${\bf r}\in S=
\{{\bf r}|{\bf r}\notin T, {\bf r}'\in T, |{\bf r}'-{\bf r}|=1\}$
if the local O$_2$ concentration is sufficiently large:
$T\to T\cup\{{\bf r}\}$ with probability 
$\Delta t/t_{\rm TC}$ if $O_2({\bf r})>c_{\rm oxy}$. 

\underline{\it Vessel growth:}
New vessels (edges in G) can be grown (added to G) in regions, where
the local GF concentration is sufficiently large: 
$i)$ Randomly choose $e_1\in E$ and 
${\bf r}_1\in e_1$.  $ii)$ If there is ${\bf r}_2\in e_2$ 
for a certain $e_2\ne e_1$ so that 
${\bf r}_2-{\bf r}_1 \parallel {\bf i}$ or ${\bf j}$ or ${\bf k}$, 
$2\leq |{\bf r}_2 - {\bf r_1}|\leq \ell_{\rm max}$, 
$GF({\bf r}({\bf r}_1,{\bf r}_2,\zeta))>c_{\rm gf}$ 
for all $\zeta = 0,1,\ldots, |{\bf r}_2-{\bf r}_1|$, 
and ${\bf r}({\bf r}_1,{\bf r}_2,\zeta)\notin e$ for any $e\in E$ 
with $\zeta = 1,\ldots, |{\bf r}_2-{\bf r}_1|-1$, 
then $E\to E\cup\{e_{\rm new}\}$ with 
$e_{\rm new}=\{{\bf r}|{\bf r}={\bf r}({\bf r}_1,{\bf r}_2,\zeta), 
\zeta=0,1,2,\ldots,|{\bf r}_2-{\bf r}_1|\}$.
Repeat $i)$ and $ii)$ $\delta\, n(E)\Delta t/t_{EC}$ times, since the
number of potential sprouting events should be proportional to the
total length of the vessels in the network.

\underline{\it Vessel dilatation:} 
Within the tumor no new vessels can occur because of the lack of
space, but vessels increase their diameter due to proliferation of EC's
in the vessel walls if the local GF concentration is sufficiently
large: For all $e\in E$, $d(e)\to d(e)+\sum_{{\bf r}\in e} \theta_{\rm
step}(GF({\bf r})-c_{\rm gf}) /(\ell(e)+1)/2\pi$ with probability
$\Delta t/t_{EC}$ if $d(e)<d_{\rm max}$. Here $\theta_{\rm step}(x)=1$
for $x\geq 0$ and $\theta_{\rm step}(x)=0$ for $x<0$.

\underline{\it Vessel regression and collapse:} 
Weakly perfused vessels can collapse inside the tumor due to the solid
stress exerted by the tumor: Compute $P({\bf r})$ for all ${\bf r}\in
V$ and then $f(e)$ and $q(e)$ according to
Eq.~(\ref{eq:poiseuille}). Vessels that are cut from the blood
circulation ($q(e)=0$) are instantaneously removed. For all other
$e\in E$ we set $E\to E-\{e\}$ with probability $\Delta t/t_{\rm
collapse}$ if $f(e)/f_0<\eta_{\rm c}$ (i.e a weak shear force), and
$n(\{{\bf r}|{\bf r}\in T, |{\bf r}-{\bf r}'|\leq 1
\ {\rm for} \ {\bf r}'\in e\})\geq 0.8\, \ell(e)$ (i.e. a large density 
of TCs around the vessel).

\underline{\it TC death:} 
TC's that are under-oxygenated longer than a time $t_{\rm max}$ will
die: For all ${\bf r}\in T$, $t_{\rm uo}({\bf r})\to t_{\rm uo}({\bf
r})+1$ if $O_2({\bf r})\leq c_{\rm oxy}$. If $t_{\rm uo}({\bf
r})>t_{\rm max}$, $T\to T-\{{\bf r}\}$ with probability $1/2$.

\begin{figure}
\center
\includegraphics[width=\columnwidth]{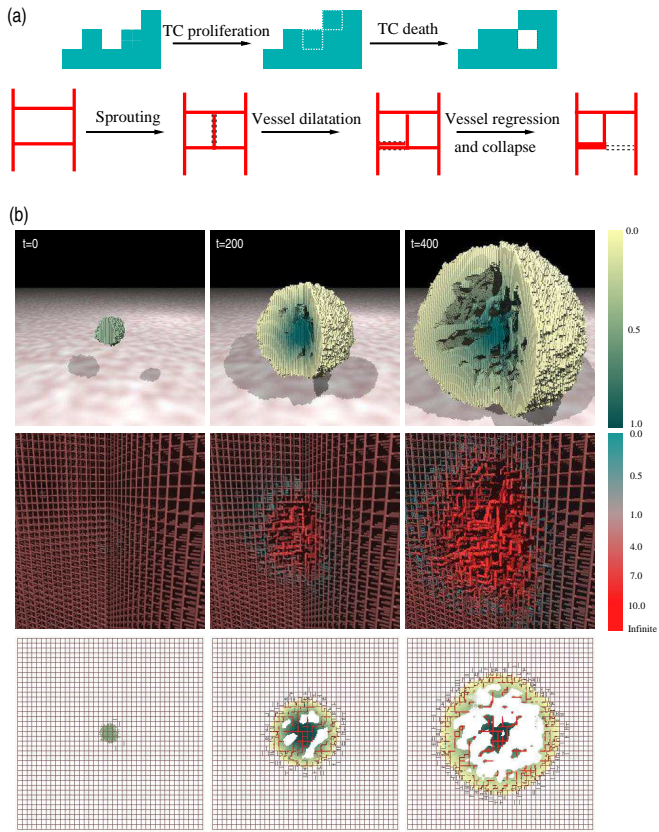}
\caption{{\bf (a)} Schematic illustration of the presented model: 
  The upper panel indicates TC proliferation and death, the lower
  panel shows vessel formation, dilatation and regression/collpase
  (the modified part of the configuration is always indicated by
  broken lines).  {\bf (b)} The time evolution of the tumor-vessel
  system is demonstrated by 3 snapshots at time t=0, 200, and 400. The
  upper panel shows only the tumor (note the necrotic regions inside),
  the middle panel only the vessel network (note the increased MVD at
  the tumor periphery, and the reduced MVD and dilated vessels in the
  tumor center), and the lower panel shows an equatorial cross section
  of the whole system in the xy plane at $z=L/2$ (lower). The
  parameters are chosen as mentioned in the text. The color code of
  the TCs represents the age scaled to $[0,1]$ and the color code of
  the vessel indicate the scaled blood flow, $q(e)/q_0$.}
\label{fig:model}
\end{figure}

A schematic illustration of these procedures is shown in
Fig.~\ref{fig:model}. We have simulated the model using various
parameter values, but here we restrict ourselves to the discussion of one
typical parameter set, which is partly guided by data for human
melanoma \cite{paku}. The original MVD is set by $\delta=100\mu m$ and
the oxygen diffusion radius $R_{\rm oxy}=100\mu m$, yielding an
average O$_2$ concentration of $\overline{O_2}\approx0.03$. The oxygen
threshold for TC proliferation and for cell death due to hypoxia is
$c_{\rm oxy}=0.01$, i.e. clearly below $\overline{O_2}$. TC and EC
proliferation time are $t_{\rm TC}=10$ h and $t_{\rm EC}=40$ h,
respectively, maximum TC survival time in hypoxia $t_{\rm max}=20$ h.
The GF diffusion radius is $R_{\rm gf}=200\mu m$, the GF threshold
$c_{\rm gf}=0.001$. The maximum vessel diameter is $d_{\rm max}=35\mu
m$, maximum sprout migration distance $\ell_{\rm max}=100\mu m$,
critical shear force $\eta_{\rm c}=0.5$ and vessel collapse rate
$1/t_{\rm collapse} = 1/50$ h. The initial tumor size
$N_{\rm TC}=27000$ (i.e. an initial tumor diameter of ca. 0.6mm).

An example for the time evolution of the tumor/vessel system in this
model is shown in Fig.~\ref{fig:model}. Starting from a regular vessel
network the MVD in the peritumoral region is increased due to the
supply of GFs from the tumor, as can best be seen in the snapshots of
an equatorial cross section through the tumor center (last row of
Fig.1). Once the tumor grows over this highly vascularized region,
vessels start to collapse, by which the MVD in the interior of the
tumor is continuously decreased until only a few thick vessels,
surrounded by cuffs of TCs remain. Due to the reduced MVD in the tumor
center regions become hypoxic and TCs will die leaving large
necrotic regions. This compartmentalization of the tumor into
different shells that can be discriminated by MVD, vessel diameter and
necrosis is also observed in real tumors \cite{paku}.

Fig.1 shows that the original vascular network, consisting of
capillaries of equal diameter arranged in a regular grid with a given
MVD that guarantees homogeneous distribution of $O_2$ and a constant
shear stress in all vessels, is dynamically transformed into a
compartmentalized network with irregularly arranged dilated vessels
and characterized by an inhomogeneous MVD and $O_2$ distribution.
This remodeling is strongly correlated with the blood flow pattern:
When new vessels are generated, they share the blood flow with their
parents, which causes all of them to have weaker shear forces, subject
to potential vessel collapse. On the other hand, when such critical
vessels are indeed removed, the blood flow has again to be redirected
into the remaining vessels, leading to an increase of the shear forces
in the involved vessels.  At the same time, surviving vessels may
increase their diameters, also resulting in higher shear forces. The
ratio $t_{\rm EC}/t_{\rm collapse}$ basically controls this
flow-correlated remodeling process consisting of
generation / dilatation and collapse and also affects the tumor volume
or necrosis: Necrosis dominates the tumor tissue in lack of oxygen
supply for a larger value of the ratio while the total necrotic region
shrinks for a smaller value.

\begin{figure}
\includegraphics[width=\columnwidth]{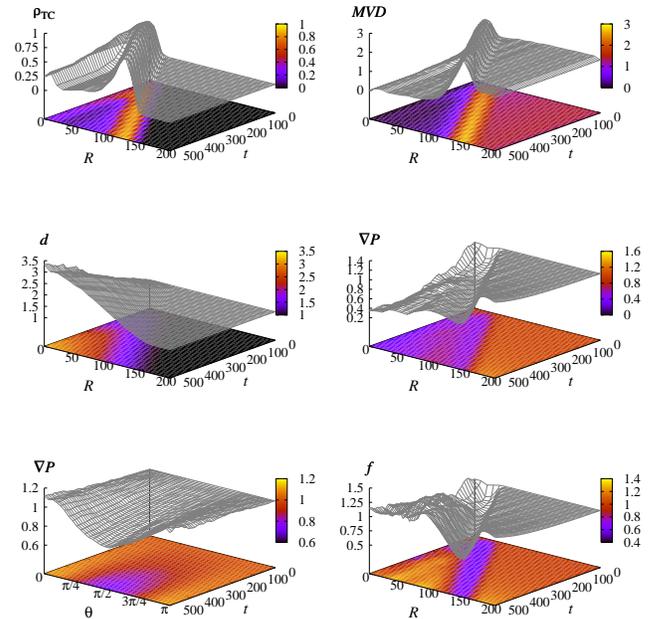}
\caption{Plots of the tumor density $\rho_{\rm TC}$, MVD, vessel diameter $d$, 
  pressure gradient $\nabla P$, and shear force $f$ as functions of
  the distance to the center $R=|{\bf r}-{\bf r}_{\rm c}|$ and time
  $t$. The lower left plot shows the dependence of the pressure
  gradient on the azimuthal angle $\theta$ between ${\bf r}-{\bf
  r}_{\rm c}$ and the diagonal $-{\bf r}_{\rm c}$.}
\label{fig:data}
\end{figure}

The dynamical evolution described thus far can be analyzed
quantitatively by studying the following quantities: The radial tumor
density $\rho_{\rm TC}(R) = n(T_R)/n(Z_R)$, vessel density
$MVD(R)=n(E_R)/(4\pi R^2)$, vessel diameter $d(R)=\sum_{e\in E_R}
d(e)/\sum_{e\in E_R}$, pressure gradient $\nabla P(R)=\sum_{e\in E_R}
\nabla P(e)/\sum_{e\in E_R}$, and $\nabla P(\theta)=\sum_{e\in
E_\theta}
\nabla P(e)/\sum_{e\in E_\theta}$, and shear force $f(R)=\sum_{e\in E_R}
f(e)/\sum_{e\in E_R}$. Her we have used $Z_R = \{{\bf r}|R\leq |{\bf
r}-{\bf r}_{\rm c}|<R+\Delta R\}$, $E_R = \{e\in E| R\leq |({\bf
r}_s(e)+{\bf r}_t(e))/2 - {\bf r}_{\rm c}|< R+\Delta R\}$, $E_\theta =
\{e\in E| \theta\leq
\cos^{-1}\frac{({\bf r}_{\rm c}-{\bf r})\cdot {\bf r}_{\rm c}}
{|{\bf r}_{\rm c}-{\bf r}||{\bf r}_{\rm c}|}< \theta+\Delta \theta\}$,
and $T_R = T\cap Z_R$.

As seen in Fig.~\ref{fig:data}, the peak of $MVD(R)$ is in accordance
with the tumor boundary $R_{\rm TC}(t)$. The tumor grows approximately
spherically with $R_{\rm TC}(t)-R_{\rm TC}(0)\simeq 2t/t_{\rm TC}$,
where the factor $2$ is typical for the Eden growth.  In the tumor
center, $MVD(R)$ and $\rho_{\rm TC}(R)$ are both very low. Their shape
similarity is due to continuous interactions between the tumor and the
vessel network. Vessels that have long been exposed to GF produced by
TCs have large diameters: $d(R)$ increases linearly from $1$ at
$R\simeq R_{\rm TC} + R_{\rm gf}$ to $d_{\rm max}$ at the tumor
center. Such a characteristic vessel morphology is also in a quantitative
agreement with experimental data from the human melanoma~\cite{paku}.

The blood pressure gradient in the tumor center is up to 50\% lower
than in normal vessels which is, from hydrodynamic considerations, an
immediate consequence of the increase MVD in the peritumoral
region. Moreover, the pressure gradient is lowest in the direction
orthogonal to the global flow $(\theta=\pi/2)$, since for a given
network the flow tends to use shortest paths from source to
sink. Finally, the shear force acting on each vessel wall depends on
the vessel diameter and the pressure gradient. The sharp decrease of
$\nabla P$ in the peritumoral region is inherited to $f(R)$ while the
increased vessel diameter in the tumor center gives rise to large
shear forces in large vessels.

\begin{figure}
\includegraphics[width=\columnwidth]{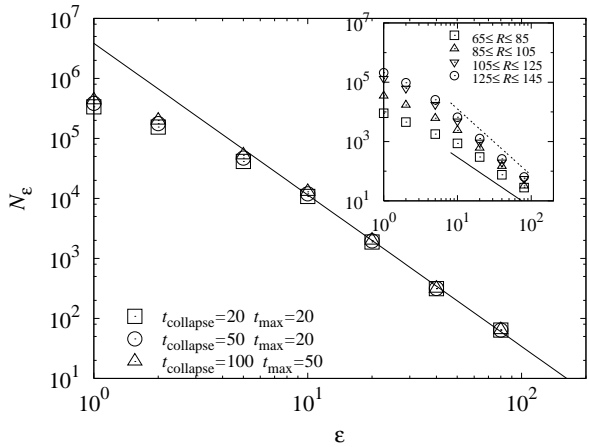}
\caption{Plots of $N_\epsilon$ of Eq.~(\ref{eq:fractal}) in the vessel 
  networks at $t=400$ for different values of $t_{\rm collapse}$ and 
  $t_{\rm max}$ with other parameters taking the typical values. 
  The tumor and the peritumoral region extend up to 
  $R\lesssim 145$, for which $N_\epsilon$ is computed. 
  In all cases, the data are fitted as $N_{\epsilon}\sim 
  \epsilon^{-2.52(5)}$. The solid line has a slope $-2.52$. 
  The inset shows $N_\epsilon$ measured in different shells of 
  the same thickness $20$ for $t_{\rm collapse}=20$ and $t_{\rm max}=20$. 
  The slopes of the upper dashed line and lower solid line are 
  $-2.24$ and $-1.68$, respectively.}
\label{fig:fractal}
\end{figure}

The geometrical features of the emerging tumor vasculature in our
model are obviously very different from the original, regular
capillary network: It consist of a combination of dense and void
regions that might possess fractal properties. We used the
box-counting method to determine the fractal dimension $D_f$ as
\begin{equation}
D_f = - \lim_{\epsilon\to 0} \ln N_\epsilon/\ln \epsilon,
\label{eq:fractal}
\end{equation}
where $N_\epsilon$ is the number of boxes of volume $\epsilon^3$
necessary to cover the tumor vessel network, that is defined to lie
within the outer limit of the peritumoral region. 
The plot of $N_\epsilon$ versus $\epsilon$ shown in
Fig.~\ref{fig:fractal} yields $D_f=2.52(5)$, in agreement with
the value for the critical percolation cluster in the random
bond-percolation process in three dimension~\cite{random-perc}. 
We checked that we obtain the same value for a wide range of 
parameter values. 

From this observation we can conclude that the basic mechanism
responsible for the fractal properties of the tumor vasculature in our
model is the stochastic removal of vessels via vessel collapse and
regression. In conventional percolation a critical cluster only
emerges if the edge concentration is fixed to be exactly at the
percolation threshold. In our model this fine tuning is not necessary:
the dynamically evolving network drives itself into this critical
state since the removal of vessels is correlated with the blood flow.
The collapse of critical vessels stabilizes the remaining ones due to
an increase in blood flow, as shown in our quantitative analysis.

It has been suggested \cite{jain-fractal} that the
origin of the fractal architecture of tumor vasculature might be due
to an underlying invasion percolation process 
\cite{invasion-perc} due to inhomogeneities
in the growth supporting matrix. In view of the theoretical model we
have presented, which does not involve any such
matrix-inhomogeneities, we propose that it is rather the flow
correlated percolation process due to collapsing vessels inside the
tumor that determines the fractal properties of the tumor
vasculature. A commonly accepted view, at least for a large class of
tumors like melanoma, also shared by our theoretical approach, is that
neo-vascularization mainly occurs at the tumor perimeter and a drastic
reduction of vessel density occurs in the interior of the tumor. In
such a scenario it appears unlikely that the fractal properties
attained during growth in the periphery, independent of having
characteristics of invasion percolation or not, survive the random
dilution process in the tumor center. Thus a theoretical model that
does not take into account vessel collapse can possibly explain the
observed fractal dimension of the vessel network in tumors where the
central MVD is {\it not} drastically reduced but fails to do so in
networks where it is.

To conclude we have introduced a theoretical model for a dynamically
evolving, three-dimensional vessel network interacting with a growing
tumor, which is guided by experimental data for human melanoma. The
emerging network morphology agrees well with those data and we find
that the network is remodeled from a regular into a fractal structure
with characteristics of random percolation. This suggests also for a
large class of real solid tumor with decreased central MVD that the
basic mechanism leading to the fractal features of the
tumor vasculature is the mechanism of stochastic vessel collapse
inside the tumor.

\end{document}